# Diamond nanoparticles as photoluminescent nanoprobes for biology and near-field optics[1]

Aurélien CUCHE [a], Yannick SONNEFRAUD [a], Orestis FAKLARIS [b], Damien GARROT [b], Jean-Paul BOUDOU [c], Thierry SAUVAGE [d], Jean-François ROCH [b], François TREUSSART [b,*], Serge HUANT [a]

[a] Institut Néel, CNRS/Université Joseph Fourier, Grenoble, France

[b] Laboratoire de Photonique Quantique et Moléculaire, UMR 8537 CNRS/Ecole Normale Supérieure de Cachan, Cachan, France

[c] BioEmCo, UMR 7618 CNRS/Université Pierre et Marie Curie, Paris, France

[d] Laboratoire Conditions Extrêmes et Matériaux : Haute Température et Irradiation, UPR 3079 CNRS, Orléans, France

* Corresponding author: treussar@physique.ens-cachan.fr

---

[1] Keynote talk by F. Treussart at the 15th International Conference on Luminescence and Optical Spectroscopy of Condensed Matter (ICL'08), Lyon, July 2008. Proceedings to appear in J. Luminescence

**ABSTRACT**: We present results showing the potential of diamond nanoparticles with size < 50 nm as photoluminescent nanoprobes for serving as stable point-like emitters attached at the tip apex of a near-field optical microscope to achieve enhanced spatial resolution.

## I. Introduction

Thanks to their unique long-term photostability and cytocompatibility, fluorescent diamond particles of size below 50 nm [1], hereafter referred to as *nanodiamonds*, have a great potential in biological applications such as marking or tracking [2, 3, 4], and in quantum optics and nano-optics, where they can be used, *e.g.*, as single-photon emitters [5] or non-blinking and non-bleaching nanoscaled light emitters [6, 7, 8]. In the present paper, we first briefly update our previous studies devoted to the use of photoluminescent nanodiamonds as markers for biology [4] and to the detection of the smallest fluorescent nanodiamonds reported so far, *i.e.* with sizes of the order of 25 nm [8]. Then we describe a new study devoted to the near-field scanning optical microscopy (NSOM) imaging and tip-induced manipulation of more heavily colour-centre doped nanodiamonds than those studied previously. We show that during scanning by the dielectric optical tip of a sample consisting in nanodiamonds dispersed on a fused silica substrate, a few of them are progressively trapped at the tip apex. We further demonstrate that these nanodiamonds remain stuck at the tip apex for a sufficiently long time to permit the acquisition of additional NSOM images and of optical spectra of the active tip which confirm, in turn, that nanodiamonds have been successfully trapped. Routes toward the trapping of a *single* nanodiamond are sketched.

## II. Previous results and experimental details

As bought type 1b diamond nanocrystals (used in polishing industry) is irradiated by a high energy and dose proton beam (2.5 MeV, dose $5.10^{15} H^+/cm^2$) to create vacancies (V) that migrate during an annealing process stabilizing them next to nitrogen (N) substitutional impurities. This process leads to the formation of NV colour centres that have a *perfect photostability* (no bleaching, no blinking) [1]. Stable water colloidal suspension of these diamond nanocrystals is obtained after cleaning annealed powder in acid baths.

We have observed by scanning confocal microscopy that nanodiamonds naturally penetrate inside cell in culture (cancerous cell of the *HeLa* line) and can be observed at the single particle and single colour-centre level [4]. Time-resolved imaging using a ps-pulsed laser as the excitation source allowed us to enhance the signal to background (due to cell autofluorescence) ratio by one order of magnitude, making tracking of a nanoparticle inside cell cytoplasm much easier in the future. We also studied the nanodiamond into cell internalization mechanism and observed that, after 2 hours of incubation, they only partly (about 20%) co-localize with endosome. Other internalization paths (lysosome in particular) are under investigation.

Using a home-made NSOM with uncoated optical tips and intensity time-correlation measurements, we observed that 30% of the nanoparticles from the colloidal suspension deposited on the silica substrate contained at least one photostable NV colour centre down to the size of 25 nm, and that almost all the nanoparticles have a size < 35 nm [8]. Here, we apply this NSOM setup (see Ref. 8 and 9 for a detailed description: the tip is used in the excitation mode, the light collection is made in transmission by a high NA objective), though without the intensity time-correlation, to a new nanodiamond specimen of the same nominal size (< 50 nm) and exposed to a larger irradiation rate (2.4 MeV, dose $5.10^{16} H^+/cm^2$). The next section focuses on the results obtained with this new specimen, in particular on the manipulation of nanodiamonds by the scanning tip.

**III. NSOM manipulation of highly irradiated nanodiamonds**

Figure 1 shows a NSOM fluorescence image and a topographic image of nanodiamonds dispersed by spin coating on a fused silica substrate that has been previously cleaned (successive ultrasonic baths in special soap, acetone and ethanol) to remove most of the surface contamination. Both types of images are acquired simultaneously thanks to a quartz tuning fork on which the tip is glued to monitor the tip-sample distance by shear-force feedback [10]. The amplitude of the shear-force is set to approximately 1 nN in the present experiments [10]. Although the topographic image of Fig. 1 is very pixelised, a careful analysis of higher resolution and shorter range images (not shown) reveals that most of the diamond particles are fluorescent and exhibit sizes in the 30 – 40 nm range as evaluated from the height of topographic features (the apparent lateral size is enlarged by the tip lateral extension). However, as small as 25 nm fluorescent nanodiamonds can also be observed (Fig. 2). It can be seen on the fluorescence image that some nanodiamonds give much weaker signal than others. This is partly due to the optical filter used in the detection branch which cuts most of the neutral $NV^0$ centre emission, blue-shifted from the $NV^-$ one [1, 8, 11]. In addition, some of the nanodiamonds might host more than one NV centre in this more heavily doped specimen as has already been observed in more lightly doped nanodiamonds [8]. It is worthwhile to notice further from Fig. 1 (and Fig. 2 as well) that the nanodiamond fluorescence spots are circular (see also ref. 8) in contrast with several reports of annular or double-lobed NSOM fluorescence spots of single molecules (see e.g. Ref.

12 and 13) or fluorescent latex nanobeads [14, 15] made with metalized optical tips. This is because metalized tips exhibit a complex polarization-dependent double-lobed [12, 16, 17] distribution of the electromagnetic near-field at the apex whereas uncoated tips should not exhibit such a complex distribution [18]. A final point that is worth commenting on in the fluorescence image of Fig. 2 is the remarkably low background signal (below 4 kcps; 1 kcps= 1000 counts per second, or in other words 1000 photons detected by the avalanche photodiode detector per second) due to the use of a pure-silica core fibre and the careful cleaning of the fused pure-silica cover slip.

After scanning successive NSOM images, we could notice that this background signal increased step wise on several lines of the scanned surface, and then decreased again. This background change is related to the displacement of individual nanodiamonds by the scanning optical tip. In some cases, it is even possible to trap an individual nanodiamond with the tip. Such a trapping translates into a persistent increase in the background signal. An example of a trapping phenomenon is shown in Fig. 2 where it can be seen on both the topographic and optical images that a 25 nm individual nanodiamond is first pushed from the left part of the images to the right part (see arrows) until it remains finally stuck on the tip. From this very moment on, the background signal increases by approximately 30 kcps which corresponds to the average signal collected from a single bright nanodiamond under the present experimental conditions.

The question that arises naturally is to know whether such an attachment is reasonably strong enough to envisage working with such a "functionalized" optical tip or whether the attached nanodiamonds will be rapidly released. To check this point, we have taken two successive optical spectra of the active tip. These are shown on Fig. 3. The first spectrum has been taken immediately after attaching the nanodiamond particles and approaching the optical tip in the near-field of the silica cover slip at a particular position free from any deposited nanodiamond. The second one has been recorded after exciting the tuning fork with a large vibration amplitude of 15 nm (instead of 5 nm in normal operation) for a few minutes, replacing the initial sample with a bare silica cover slip, and then approaching the tip into the optical near-field again (distance < 50 nm) thanks to the tuning fork feedback loop. As can be seen, both spectra can be very well superimposed which shows that the optical tip has not been modified during this manipulation. Both spectra are typical of NV colour-centres in diamond. They are a superposition of contributions from neutral $NV^0$ and negatively-charged $NV^-$ centres [1, 11]. This confirms that fluorescent nanodiamonds have actually been attached to the optical tip and remain stuck there, even after "shaking" the tip. This also indicates that either a single particle with several NV centres of different charge states or several particles have been attached. The increase in the background signal already seen prior to the single-particle attachment event of Fig. 2 gives support to the latter explanation.

**IV Conclusions and perspectives**

We have demonstrated that it is possible to attach nanoscaled (approximately 25 nm in diameter) fluorescent diamond particles at the apex of an uncoated optical tip just by scanning a rather standard sample obtained by spin-coating a nanodiamond solution on a pure-silica cover slip. The optical tip is a bare one, i.e., it is neither metal coated nor covered with a thin glue or PMMA layer, as done previously for attaching larger diamond particles [6] or CdSe quantum dots [9, 19], respectively. In the present preliminary successful attempt, the attachment method is not under full control. Such a control would require several criteria to be met simultaneously such as, *e.g.*, a rigorous control of the tip-sample shear force [10], because large forces in excess of 1 nN can modify the surface [20], and careful scans of the sample prior to attempting the attachment of a single well-selected nanodiamond. This is not far from reach. Also, one might ask oneself what is the sticking mechanism. Is it of a van-der-Walls nature only? Is there an electrostatic contribution too? If so, can it be controlled, *e.g.*, by using specially designed optical tips and/or controlling the environment conditions? Further work is needed to find answers to these questions.

**ACKNOWLEGMENTS**

This work is partly supported by the European Commission through the Nano4Drugs (contract LSHB-2005-CT-019102), and the EQUIND (contract IST-034368) projects. Aurélien Cuche acknowledges funding by the Région Rhône-Alpes through the "Cluster Micro-Nano."

**References**

[1] F. Treussart, V. Jacques, E. Wu, T. Gacoin, P. Grangier, J.-F. Roch, Physica B 376-377 (2006) 926.

[2] C.-C. Fu, H.-Y. Lee, K. Chen, T.-S. Lim, H.-Y. Wu, P.-K. Lin, P.-K. Wei, P.-H. Tsao, H.-C. Chang, W. Fann, Proc. Natl. Acad. Sci. USA 104 (2007) 727.

[3] Y.-R. Chang, H.-Y. Lee, K. Chen, C.-C. Chang, D.-S. Tsai, C.-C. Fu, T.-S. Lim, Y.-K. Tzeng, C.-Y. Fang, C.-C. Han, H.-C. Chang, W. Fann, Nature Nanotech. 3 (2008) 284.

[4] O. Faklaris, D. Garrot, V. Joshi, F. Druon, J.-P. Boudou, Th. Sauvage, P. Georges, P.Curmi, F. Treussart, to appear in Small (2008) ; O. Faklaris, A. Elli, V. Joshi, T. Sauvage, J.-P. Boudou, J.-F. Roch, P. Curmi, F. Treussart, to appear in *MRS Fall Meeting 2007 proceedings*, Symposium P - Diamond Electronics, fundamentals to applications II (Boston, 26-30 November 2007).

[5] A. Beveratos, S. Kühn, R. Brouri, T. Gacoin, J.-P. Poizat, P. Grangier, Eur. Phys. J. D 18 (2002) 191.

[6] S. Kühn, C. Hettich, C. Schmitt, J.-P. Poizat, V. Sandoghdar, J. Microsc. 202 (2001) 2.

[7] J. R. Rabeau, A. Stacey, A. Rabeau, F. Jelezko, I. Mirza, J. Wrachtrup, S. Prawer, Nano Lett. 7 (2007) 3433.

[8] Y. Sonnefraud, A. Cuche, O. Faklaris, J.-P. Boudou, T. Sauvage, J.-F. Roch, F. Treussart, S. Huant, Opt. Lett. 33 (2008) 611.

[9] Y. Sonnefraud, N. Chevalier, J. F. Motte, S. Huant, P. Reiss, J. Bleuse, F. Chandezon, M. T. Burnett, W. Ding, S. A. Maier, Opt. Express 14 (2006) 10596.

[10] K. Karrai, R.D. Grober, Appl. Phys. Lett. 66 (1995) 1842.

[11] Y. Dumeige, F. Treussart, R. Alléaume, T. Gacoin, J.-F. Roch, P. Grangier, J. Lumin. 109 (2004) 61.

[12] E. Betzig, J.K. Trautman, T.D. Harris, J.S. Weiner, R.L. Kostelak, Science 251 (1991) 1468.

[13] N.F. van Hulst, J.-A. Veerman, M.F. García-Parajó, L. Kuipers, J. Chem. Phys. 112 (2000) 7799.

[14] C. Höppener, D. Molenda, H. Fuchs, A. Naber , Appl. Phys. Lett. 80 (2002) 1331.

[15] A. Drezet, S. Huant, J.C. Woehl, J. Lumin. 107 (2004) 176.

[16] R. S. Decca, H. D. Drew, K. L. Empson, Appl. Phys. Lett. 70 (1997) 1932.

[17] A. Drezet, M.J. Nasse, S. Huant, J.C. Woehl, Europhys. Lett. 66 (2004) 41.

[18] S. Bozhevolnyi, B. Vohnsen, J. Opt. Soc. Am. B 14 (1997) 1656.

[19] N. Chevalier, M.J. Nasse, J.C. Woehl, P. Reiss, J. Bleuse, F. Chandezon, S. Huant, Nanotechnology 16 (2005) 613.

[20] C. Obermüller, A. Deisenrieder, G. Abstreiter, K. Karrai, S. Grosse, S. Manus, J. Feldmann, H. Lipsanen, M. Sopanen, Appl. Phys. Lett. 75 (1999) 358.

**Figure captions:**

**Figure 1 :** NSOM image of nanodiamonds dispersed on a fused silica substrate. Left: topography ; right: fluorescence. Image size: 5×5 µm² for 64×64 pixels². Integration time per pixel : 80 ms. Cw laser power (wavelength: 488 nm) transmitted to the far-field: 115 µW.

**Figure 2:** Same as Fig. 1 except for the scanning area: 2×2 µm² and the laser power: 175 µW. Arrows point to the trapping event of a single nanodiamond by the optical tip. The dashed circle is intended to help visualizing two 20 nm nanodiamonds.

**Figure 3:** Two consecutive spectra of the optical tip taken after scanning the nanodiamond sample. The integration time on the CCD array detector is 180 s and the laser power at the tip apex is 12 µW in both cases. The red spectrum has been taken just after recording several images similar to that of Fig. 2 while the black one has been taken after manipulating the optical tip as described in the text.

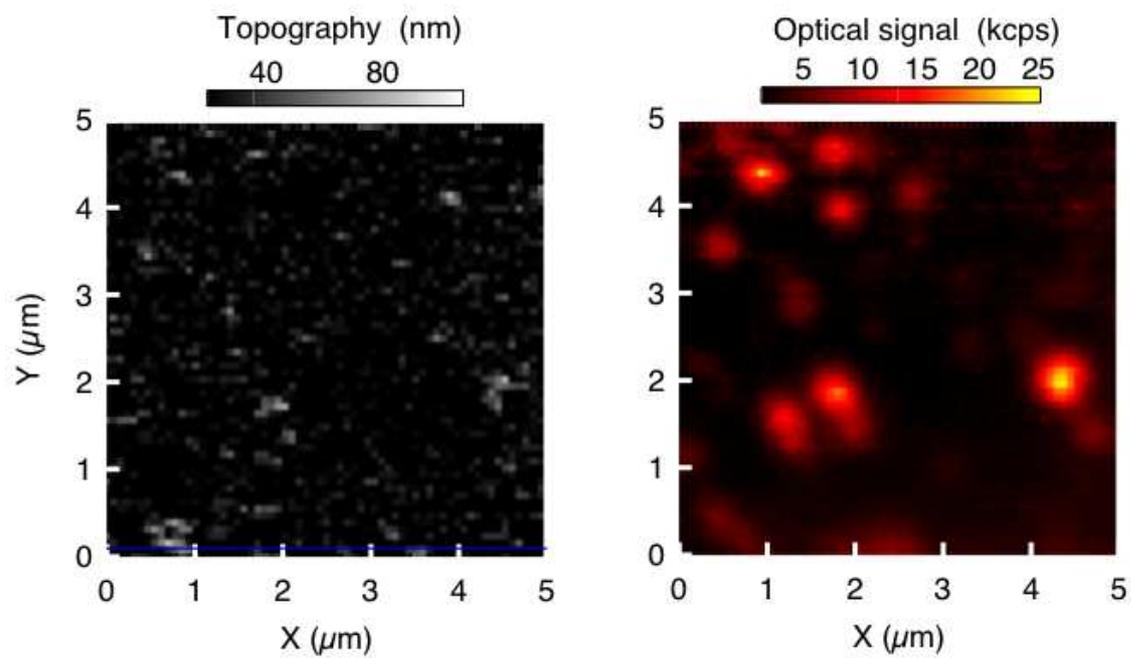

**Figure 1**

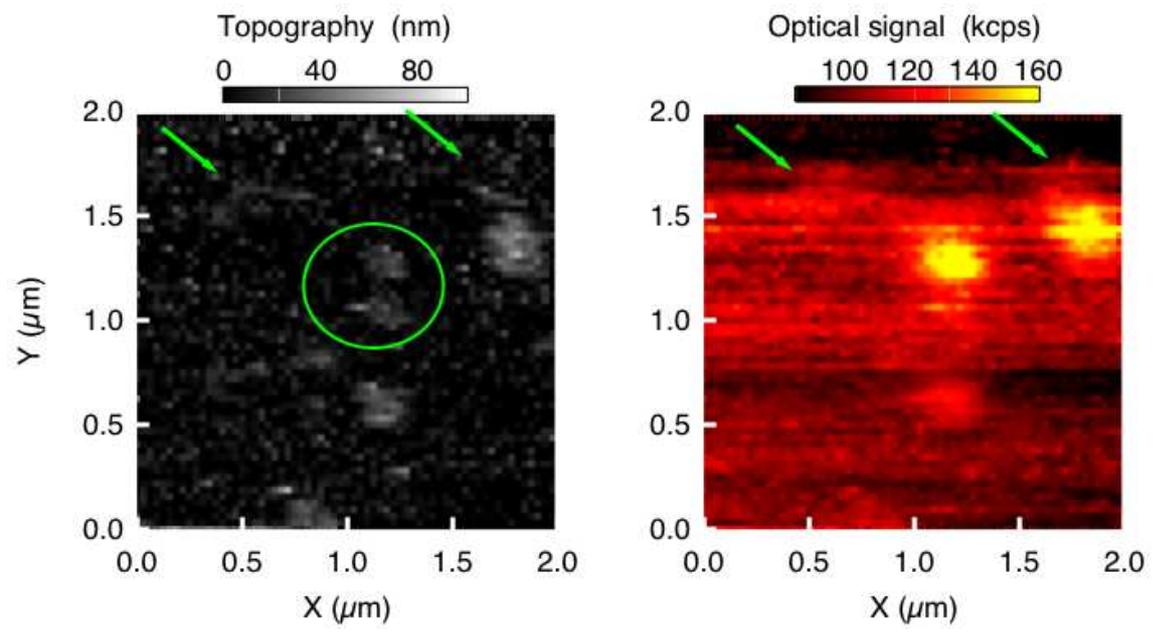

**Figure 2**

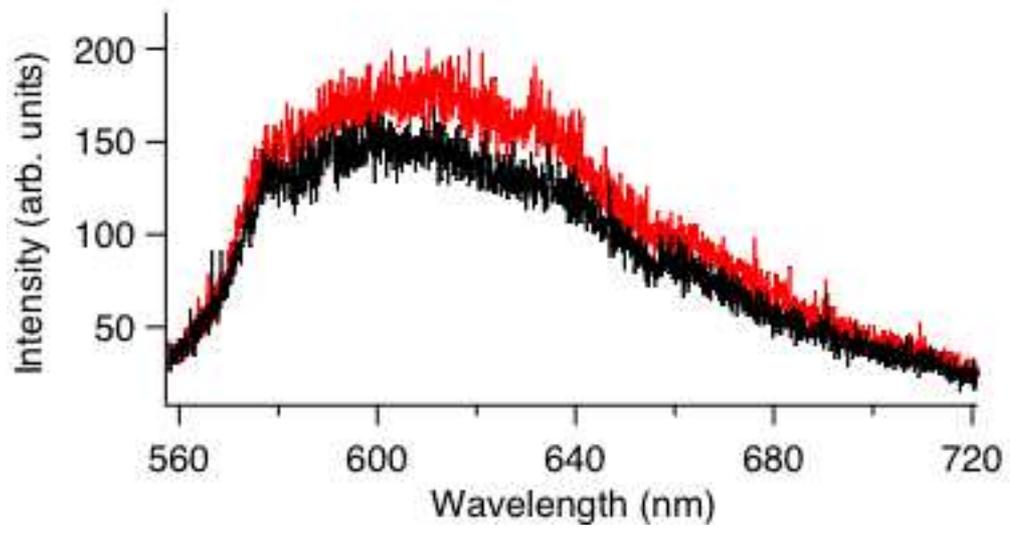

**Figure 3**